\documentstyle[prl,aps,psfig]{revtex}
\newcommand{\BEQ}{\begin{equation}}     
\newcommand{\BEA}{\begin{eqnarray}}
\newcommand{\EEQ}{\end{equation}}       
\newcommand{\EEA}{\end{eqnarray}}
\def\be{\begin{equation}}
\def\ee{\end{equation}}
\def\bc{\begin{center}}
\def\ec{\end{center}}
\newcommand{\D}{{\rm d}}

\begin{document}

\input epsf.sty
\twocolumn[\hsize\textwidth\columnwidth\hsize\csname %
@twocolumnfalse\endcsname

\draft
\widetext 

\phantom{.}
]

\narrowtext

{\bf Henkel and Pleimling reply:}
In the preceeding comment, Corberi, Lippiello and Zannetti \cite{Corb02}
studied the scaling of the linear autoresponse function
$R(t,s) = s^{-1-a} f_R(t/s)$
of ageing systems in the scaling regime of a quench below $T_c$. Here $s$
is the waiting time, $t$ the observation time, $f_R$ a scaling function and
$a$ an exponent to be determined. The exponent $a$ also describes the
scaling of the thermoremanent magnetization
\BEQ \label{Mtrm}
M_{\rm TRM}/h = \int_{0}^{s} \!\D u \: R(t,u) \sim s^{-a} F(t/s)
\EEQ
where $h$ is the external magnetic field. 
While we had found previously that scaling is observed if $a=1/2$ is taken 
\cite{Henk01}, they
assert that the asymptotic value of $a$ should be different. Specifically,
they suggest $a=1/4$ and $a=1/2$ with logarithmic corrections, for the $2D$
and $3D$ kinetic Ising model, respectively \cite{Corb02}. This assertion
is supported by numerical data for the field-cooled magnetization
$M_{\rm ZFC}/h = \int_{s}^{t} \!\D u \: R(t,u) \sim s^{-a} G(t/s)$  
obtained for both the $2D$ and $3D$ Ising model. In addition, they argue that 
for $M_{\rm TRM}$, the true asymptotic scaling behaviour should set in
only for very large values of $s$. Instead, they suggest that their data
rather fall into a preasymptotic regime with an effective 
$a_{\rm eff}=\lambda/z$, where $z$ and $\lambda$ are the dynamical and
Fisher-Huse exponents \cite{Fish88}, respectively. This is supported by
numerical data going up to waiting times $s\simeq 200$ in $3D$ and 
$s \simeq 2\cdot10^3$ in $2D$. The truly asymptotic regime is not yet reached 
by their data, but they expect to recover the same value of $a$ as obtained 
from $M_{\rm ZFC}$ \cite{Corb02}. 

\begin{figure}
\centerline{\epsfxsize=2.7in\epsfbox
{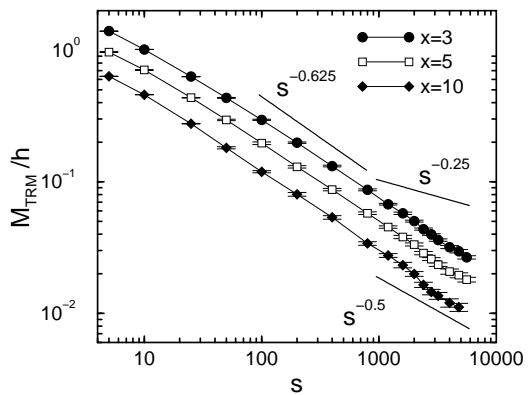}
}
\caption{Scaling of $M_{\rm TRM}$ as a function of $s$ for the zero-temperature
$2D$ Ising model.
\label{Abb1}} 
\end{figure}

\begin{figure}
\centerline{\epsfxsize=2.5in\epsfbox
{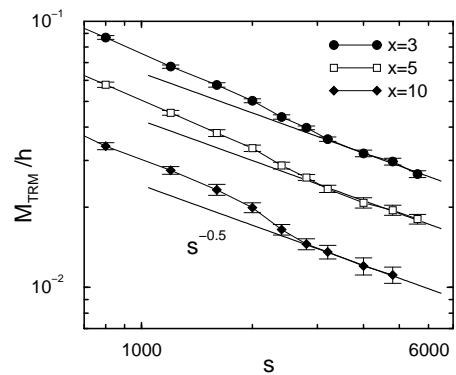}
}
\caption{Power-law scaling of $M_{\rm TRM}$ for large values of $s$.
\label{Abb2}} 
\end{figure}

In order to test the proposition of \cite{Corb02}, we obtained data for 
$M_{\rm TRM}$ for the $2D$ Ising model at zero temperature, for very long 
waiting times up to $s=5600$. In analysing the scaling of $M_{\rm TRM}$, we did
not perform any subtraction in order to isolate an `ageing part', in
contrast to the procedure adopted in \cite{Corb02}. Our result is shown in
Figure~\ref{Abb1}, for several values of the scaling variable $x=t/s$. 
To guide the eye, we also show the power laws $s^{-1/4}$, $s^{-\lambda/z}$
($\lambda/z=0.625$ in $2D$ \cite{Fish88}) and $s^{-1/2}$. In figure~\ref{Abb2}
we focus on the scaling behaviour for large values of $s$. 

At first sight, the data from Figure~\ref{Abb1} appear to be roughly consistent 
with a simple power law (\ref{Mtrm}), with $a$ between $0.5$ and $0.625$. 
On closer inspection, three regimes are visible. 
In the first regime, for small waiting times,
an approximate scaling with an effective $a_{\rm eff}\simeq 1/2$ is obtained.
A second well-defined scaling regime exists for intermediate times
(here approximately between $10 \leq s \leq 10^3$) and 
$M_{\rm TRM}\sim s^{-0.625}$. Finally, for very long times, there is a 
cross-over into a third regime, where $M_{\rm TRM}\sim s^{-1/2}$, see
figure~\ref{Abb2}. 
The same scaling regimes can be found for any temperature $T<T_c$ \cite{Henk03},
as expected, see \cite{Bray94}, but data for $T>0$ are 
much more difficult to obtain. 
We do not find any hint for a regime with $a=1/4$. In \cite{Corb02}, working
with smaller values of $s$, only the first two regimes were observed. 

We also point out that the functional form of $f_R(x)$ 
(and consequently $F(x)$) is independent of $s$. 

In conclusion, we find evidence that the truly asymptotic scaling of the
thermoremanent magnetization should be described by an exponent $a=1/2$ in the 
$2D$ Ising model. This is different from the value suggested from the analysis 
of $M_{\rm ZFC}$ \cite{Corb02}. 
This work was supported by CINES Montpellier (projet pmn2095).\\ 

\noindent M. Henkel$^a$ and M. Pleimling$^b$\\
$^a$Laboratoire de Physique des Mat\'eriaux CNRS UMR 7556, 
Universit\'e Henri Poincar\'e Nancy I,
F -- 54506 Vand{\oe}uvre les Nancy, France\\
$^b$Institut f\"ur Theoretische Physik I, Universit\"at Erlangen-N\"urnberg,
D -- 91058 Erlangen, Germany


\vspace{-10mm}

\end{document}